\documentclass[a4paper]{jpconf}
\usepackage{graphicx}
\usepackage{iopams}
\newcommand{\pom}{{I\!\!P}}
\newcommand{\reg}{{I\!\!R}}

\newcommand{\mx}{M_{_{\rm X}}}
\newcommand{\my}{M_{_{\rm Y}}}

\begin{document}
\title{QCD fits in diffractive DIS revisited}

\author{F.\ A.\ Ceccopieri$^1$, L.\ Favart$^2$}
\address{$^1$ IFPA, Universit\'e de Li\`ege,  All\'ee du 6 ao\^ut, B\^at B5a, 4000 Li\`ege, Belgium.}
\address{$^2$ Universit\'e Libre de Bruxelles, Boulevard du Triomphe, 1050 Bruxelles, Belgium.}

\begin{abstract}
A new method of extracting diffractive parton distributions is presented which avoids 
the use of Regge theory ansatz and is in much closer relation with the factorization theorem 
for diffarctive hard processes.
\end{abstract}

\section{Introduction}
\noindent\\
Diffractive parton distributions functions (DPDF) are essential ingredients in the understanding and 
description of hard diffractive processes. 
The factorization theorem for diffractive process in Deep Inelastic Scattering (DIS)~\cite{DDISfactorization,GTV} enables one to factorize the diffractive DIS cross-sections into the 
long-distance contribution parametrised by DPDF's, from the short-distance, pertubatively calculable, one. 
Although DPDF's encode non-perturbative effects of QCD dynamics,  
their variation with respect to the factorization scale is predicted by pQCD~\cite{TV,extM}.  
Moreover the short distance cross-sections is the same as inclusive DIS~\cite{DDISfactorization} 
so that higher order corrections can be systematically evaluated.
Due to the factorization theorem, DPDF's are predicted to be universal distributions
in the context of diffractive DIS. Next-to-leading order predictions based on DPDF's have been found to well describe DIS diffractive dijet cross-sections~\cite{H1LRGdiffdijet,ZEUSdifdijet},
thus confirming factorization. 
On the contrary, the issue of factorization is still debated in
diffractive photoproduction of dijets
since the H1~\cite{ddijet_photo_H1} and ZEUS~\cite{ddijet_photo_ZEUS} collaborations
do report conflicting results. 

\noindent DPDF's give the joint probability for a  
parton in a proton of four-momentum $P$ to initiate a hard
scattering keeping the proton intact with a four-momentum $P'$ in the
final state. 
Diffractive parton distributions are a special case of extended fracture functions \cite{extM} in the kinematic range $x_\pom \leq 10^{-2}$ and $|t| \leq 1 \;\mbox{GeV}^2$, where  $x_\pom$ is the fractional energy loss of the final state proton and $t$ the invariant momentum transfer, $t=(P-P')^2$.
The presence of large and positive scale violations up to the largest 
parton fractional momentum, $\beta$, accessible to experiments 
reveals that DPDF's are gluon dominated distributions. 
Diffractive gluon distributions are however hardly determined by the diffractive structure functions 
alone and this fact has stimulated experimental collaborations to measure quantity that are directly 
sensitive to it~\cite{H1LRGdiffdijet,ZEUSdifdijet,H1FLD} and to include such data in global fits. 

\noindent The general method used to extract DPDF's from available data heavily relies on a number 
of assumptions motivated by Regge phenomenology.  The purpose of this short note 
is to present a new method which does not require any Regge assumptions and is much 
closer in spirit to the factorization theorem. 

\section{Data set and observable}
\noindent\\
For this analysis we will use H1~\cite{H1LRG06} data in which DIS diffractive events, 
$ep \rightarrow eXY$, are selected by requiring a large rapidity gap 
between the hadronic system X and the low mass dissociative system Y 
(which may consist in the scattered $p$ only).
The kinematics of diffractive events is specified by the following variables:
\begin{equation}
\beta=\frac{Q^2}{Q^2+\mx^2}; \;\; x_\pom = \frac{x_B}{\beta}; \;\; y=\frac{Q^2}{s x_B}\,,
\end{equation}
where $Q^2$ is the photon virtuality, $\mx$ is the invariant mass of the
hadronic final state $X$, $x_B$ is the usual $x$-Bjorken DIS
variable 
and $\sqrt{s}=318 \; \mbox{GeV}$ is the HERA center of mass energy.  
The measurament is integrated over the hadronic final state $Y$ mass 
region $\my<1.6 \, \mbox{GeV}$ and the invariant momentum transfer
$|t|< 1 \, \mbox{GeV}^2$, which 
then define the extracted DPDF's~\cite{CT}.
The kinematical coverage is wide ranging from $3.5 \le Q^2 \le 1600 \mbox{GeV}^2$, 
$3\cdot 10^{-4} \le x_\pom \le 3\cdot 10^{-2}$ and $10^{-3} \le \beta \le 0.8$.
We notice however that the $\{\beta,Q^2\}$ coverage changes with $x_\pom$ 
due to kinematic constraints and this fact will reflect on DPDF's extraction. 
Data are presented as a three-fold reduced $e^+p$ cross section 
which depends on the diffractive structure functions 
$F_2^{D(3)}$ and $F_L^{D(3)}$. In the one-photon exchange approximation, it reads:
\begin{equation}
\sigma_r^{D(3)}(\beta,Q^2,x_\pom)=F_2^{D(3)}(\beta,Q^2,x_\pom)-\frac{y^2}{1+(1-y)^2} F_L^{D(3)}(\beta,Q^2,x_\pom)\,.
\end{equation}

\section{The new method}
\noindent\\
The widely used approach~\cite{H1LRGdiffdijet,H1LRG06,Zeus_combo,recent_global_fit} 
to extract DPDF's is to assume proton vertex factorization, \textsl{i.e.} that DPDF's can be factorized into a flux factor depending only on $x_\pom$ and $t$ and a term depending only 
on $\beta$ and $Q^2$: 
\begin{equation}
{F}_i^D (\beta, Q^2 , x_\pom , t) = f_{\pom/P} (x_\pom , t) \; {F}_i^{\pom} (\beta, Q^2 ) 
+ f_{\reg/P} (x_\pom , t) \; {F}_i^{\reg} (\beta, Q^2 ) + ...
\end{equation}
Each term in the expansion, according to Regge theory, is supposed to give a dominant contribution in a given range of $x_\pom$, the pomeron ($\pom$) at low $x_\pom$ and the reggeon ($\reg$) at higher value of $x_\pom$. The flux factor  $f_{\pom/P}$ ($f_{\reg/P}$) can be interpreted as the probability that a pomeron (reggeon) with a given value of $x_\pom$ and $t$ couples to the proton. 
The fluxes $x_\pom$ and $t$ dependences used in the fits are motivated by 
Regge theory.
This approach requires that parton distributions of the pomeron, ${F}_i^{\pom}$,
and of the reggeon, ${F}_i^{\reg}$, must be simultaneously extracted from data.
This procedure introduces a large number of parameters in the fit and it is potentially biased 
by the choices of the flux factors. A common choice used in phenomenological application 
is to fix ${F}_i^{\reg}$ to be equal to pion parton distribution
functions. Although 
such an approach has been proven to be supported by phenomenological analyses
within the precision obtained with HERA-I data, it is not  
routed in perturbative QCD
and may show up to be not satisfactory with the expected precision
increase of HERA-II data and H1+ZEUS complete combination. 

\noindent
The alternative method we propose is instead inspired by the factorization theorem 
\cite{DDISfactorization} for diffractive DIS itself. 
The latter states that factorization holds at fixed 
values of $x_\pom$ and $t$ so that the parton content described by
${F}_i^{D}$ is uniquely fixed by the kinematics of the outgoing proton
and it is in principle different for different values of $x_\pom$. 
In practice this idea is realized performing a series of separate pQCD fits 
at fixed values of $x_\pom$ with a common initial condition 
controlled by a set of parameters $\{p_i\}$. 
This procedure guide us to infere the approximate dependence of parameters $\{p_i\}$ on $x_\pom$
allowing the construction of initial condition in the $\{\beta,x_\pom\}$ space
to be used in a global fit, without any further model dependent assumption.

\newpage   
\section{Fit procedure and results}
\noindent\\
In each $x_\pom$-bin for which data are presented by the experimental collaboration 
we perform a separate pQCD fits. For the singlet and gluon distributions at the arbitrary scale $Q_0^2$ we choose:
\begin{eqnarray}
\label{ic}
\beta \; \Sigma({\beta,Q_0^2}) &=& A_q \; \beta^{B_q} \; (1-\beta)^{C_q} \; e^{-\frac{0.01}{1-z}}\,,\nonumber\\
\beta \; g ({\beta,Q_0^2}) &=& A_g \; e^{-\frac{0.01}{1-z}}\,,
\end{eqnarray}
so that there are four free parameters. 
We make the common assumption that all lights quark distributions are equal to each other
and the exponential dumping exponential factor allows more freedom in the variation of the parameters $C_q$ at large $\beta$.
The functional form in eq.~(\ref{ic}) is identical to the one used by H1 collaboration in Ref.~\cite{H1LRG06}. Since diffractive DIS data hardly discriminate~\cite{H1LRG06} between different behaviour of gluon distributions at large $\beta$
we choose the simpler one in which the gluon is a costant at $Q_0^2$. 
Such distributions are then evolved with the \texttt{QCDNUM17}~\cite{QCDNUM17} program  
within a fixed flavour number scheme to next-to-leading order accuracy.
Heavy flavours contributions are taken into account in the general massive 
scheme.
The convolution engine of \texttt{QCDNUM17} is used to obtain 
$F_2^{D(3)}$ and $F_L^{D(3)}$ structure functions at next-to-leading order 
which are then minimized versus data. In order to avoid the resonance region, 
a cut on the invariant mass
of the hadronic system $X$ is applied, $\mx^2\ge 4  \, \mbox{GeV}^2$.
Single $x_\pom$-fit results are sensitive 
to the choice of the mininum value $Q^2$ of data to be included in the fits.
The inclusion of data with $Q^2 < 8.5 \, \mbox{GeV}^2$ in general worsen the 
$\chi^2$ and induce large fluctuation in the gluon distribution.
This instability has been already noticed in Ref.~\cite{H1LRG06} and avoided 
by including in the fit only data for which $Q^2 \ge 8.5 \, \mbox{GeV}^2$. 
We will adopt here the same strategy. \\
\noindent
An essential condition for the procedure to work is that good quality fits are all obtained 
with the common initial condition, eq.~(\ref{ic}). From fits results 
presented in Table \ref{results_fixed_xpom_fits} we conclude that the initial condition provided 
by eq.~(\ref{ic}) is general enough to describe data in all $x_\pom$-bins. 
\begin{table}[t]
\label{single_xpomfit_res}
\caption{ Fit results at fixed xpom. Fits are performed taking into account only statistical errors only. Only data for which $\mx^2\ge 4  \, \mbox{GeV}^2$ and $Q^2 \ge 8.5 \, \mbox{GeV}^2$ are included in the fits.}
\begin{center}
\lineup
\begin{tabular}[t]{*{2}{l}}
\br
$x_\pom$ & $\chi^2$ / d.o.f\\
\mr
0.001 & 0.921 \\
0.003 & 0.875 \\
0.01 & 0.882 \\
0.03 & 0.472 \\
\br
\label{results_fixed_xpom_fits}
\end{tabular}
\end{center}
\end{table}
The dependence of the parameters 
on $x_\pom$ is reported in Figures from 1 to 4. 
Red dots are the results from pQCD fits at fixed $x_\pom$.  
The singlet normalization $A_q$ behaves as an inverse power of $x_\pom$. In order to improve 
the description at higher $x_\pom$, however, an additional term is also included:   
\begin{equation}
\label{aq}
A_q(x_\pom) = A_{q,0} \; (x_\pom)^{A_{q,1}} \; (1-x_\pom)^{A_{q,2}} \,.
\end{equation}
The gluon normalization is compatible with a single inverse power behaviour of the type:
\begin{equation}
\label{ag}
A_g(x_\pom) = A_{g,0} \; (x_\pom)^{A_{g,1}}\,.
\end{equation}
The coefficients $B_q$ and $C_q$ which control the $\beta$-shape of the singlet distribution 
are well described by:
\begin{eqnarray}
B_q(x_\pom) &=& B_{q,0} + B_{q,1} \, x_\pom \,, \\
C_q(x_\pom) &=& C_{q,0} + C_{q,1} \, x_\pom \,. 
\label{bcq}
\end{eqnarray}
In order to facilitate the comparison, curves resulting from fits with functional form specified in eqs. (6-9) are superimposed to points in Figures 1-4. 
\begin{figure}[t]
\begin{center}
\begin{minipage}{14pc}
\includegraphics[width=14pc]{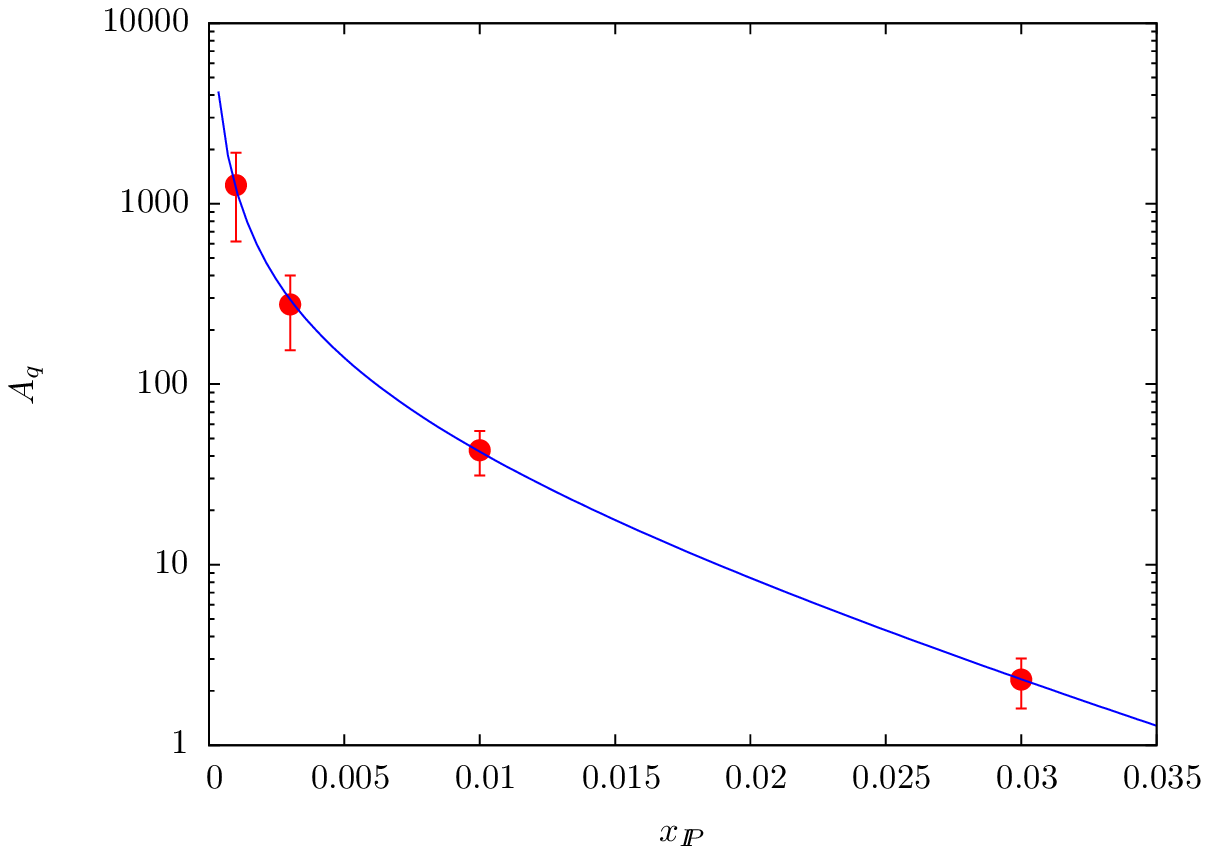}
\caption{$A_q$ as a function of $x_\pom$.}
\vspace*{0.5cm}
\end{minipage}\hspace{3pc}
\begin{minipage}{14pc}
\includegraphics[width=14pc]{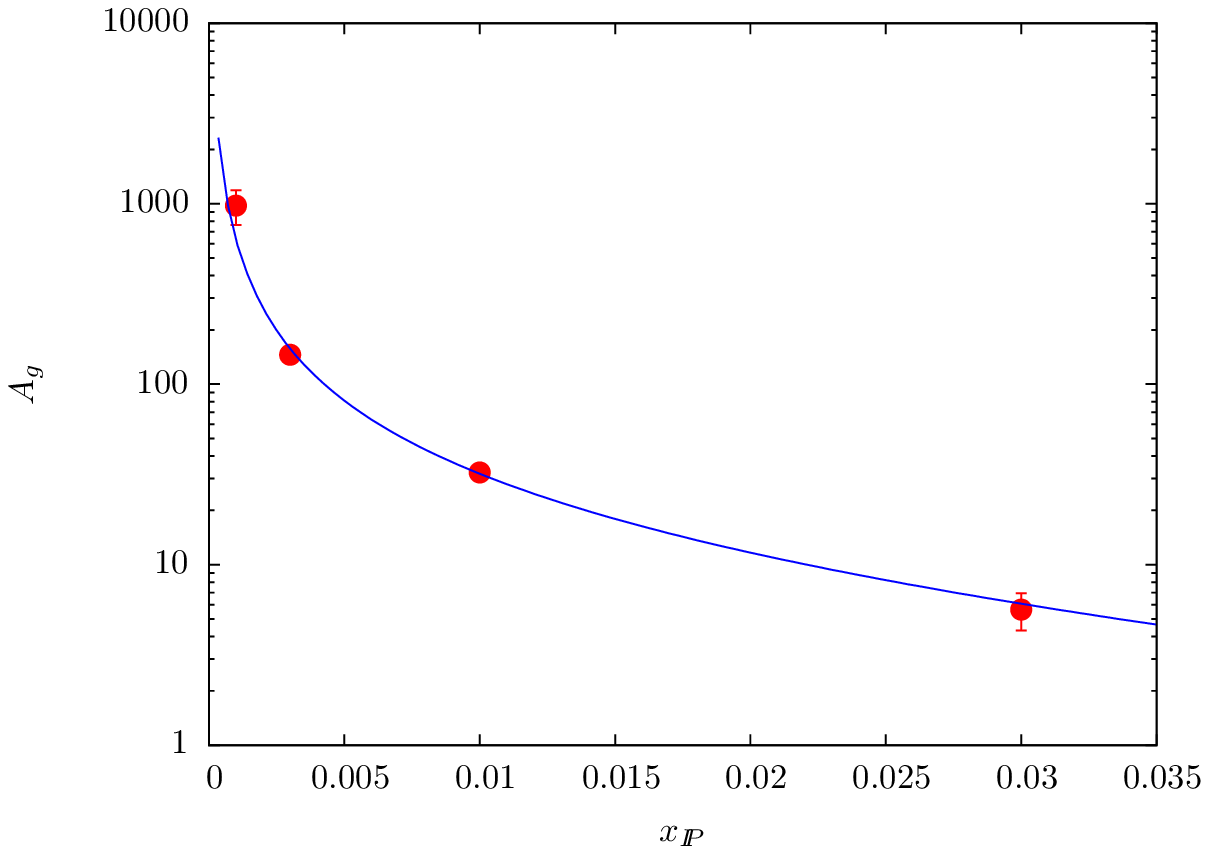}
\caption{$A_g$ as a function of $x_\pom$.}
\vspace*{0.5cm}
\end{minipage} 
\begin{minipage}{14pc}
\hspace{0.5pc}
\includegraphics[width=13pc]{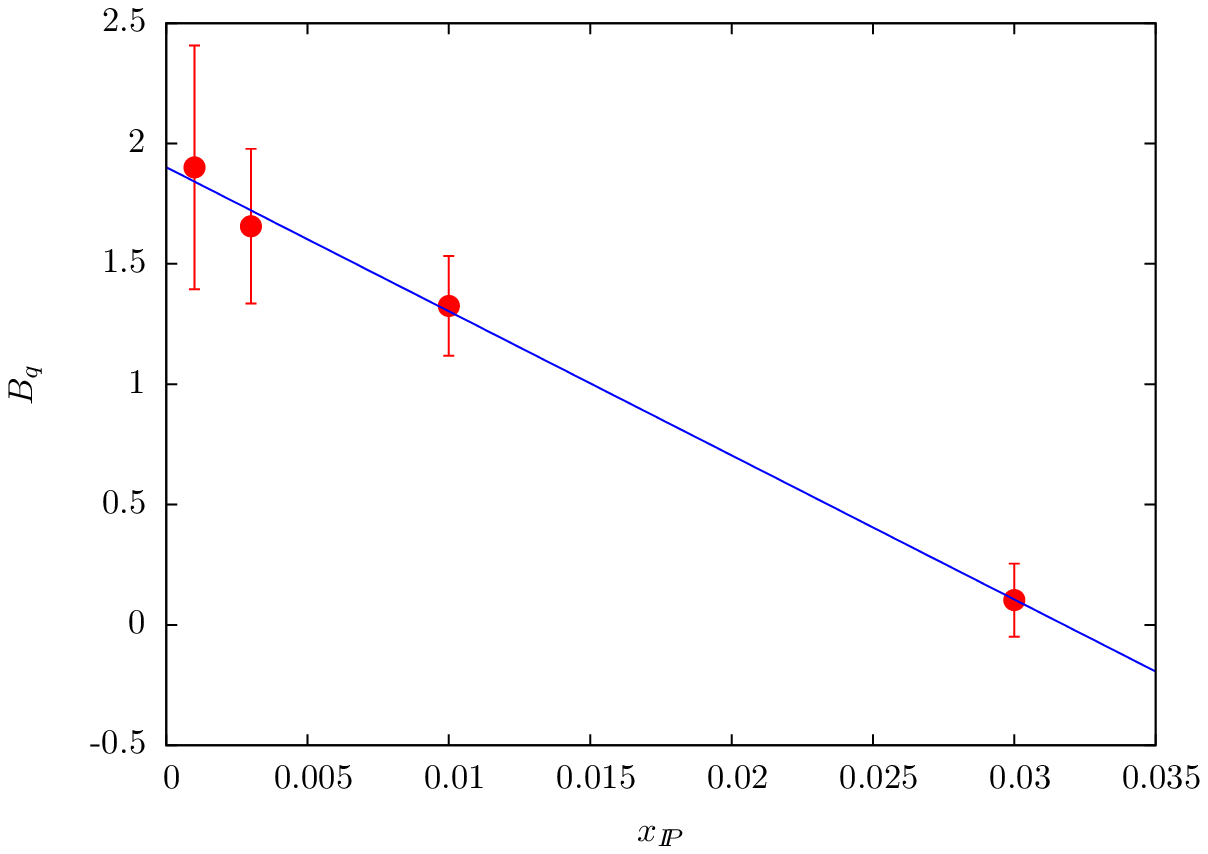}
\caption{$B_q$ as a function of $x_\pom$.}\end{minipage}\hspace{3pc}
\begin{minipage}{14pc}
\hspace{0.5pc}
\includegraphics[width=13pc]{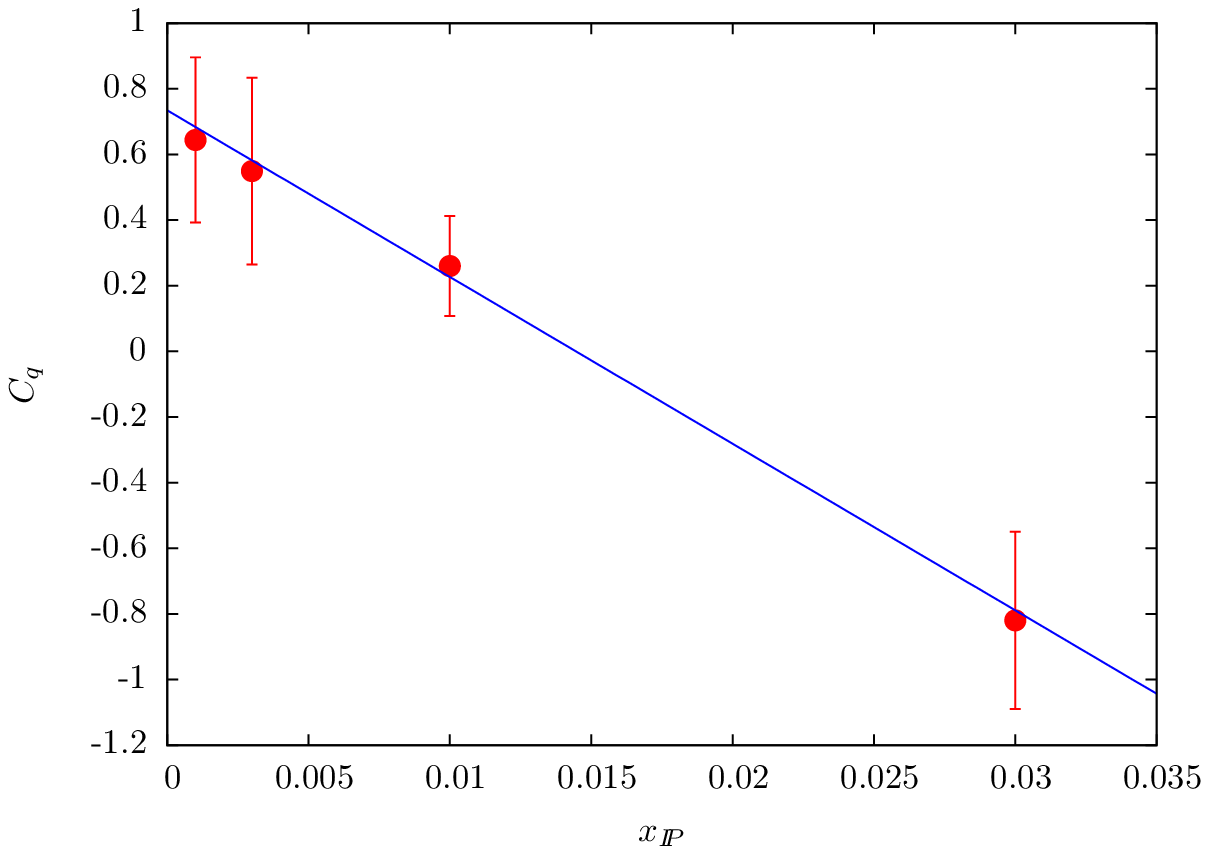}
\caption{$C_q$ as a function of $x_\pom$.}
\end{minipage} 
\vspace*{-0.5cm}
\end{center}
\end{figure}
Since the approximate behaviour of the 
parameters against $x_\pom$ is now known, we can perform a 
 $x_\pom$-bin combined fit, using the generalized initial condition 
\begin{eqnarray}
\label{ic2}
\beta \; \Sigma({\beta,Q_0^2,x_\pom}) &=& A_q(x_\pom) 
\; \beta^{B_q(x_\pom)} \; (1-\beta)^{C_q(x_\pom)} \; e^{-\frac{0.01}{1-z}}\,,\\
\beta \; g ({\beta,Q_0^2,x_\pom}) &=& A_g(x_\pom) \; e^{-\frac{0.01}{1-z}}\,,
\end{eqnarray}
with the parameters depending on $x_\pom$ as specified in eqs.~(6-9).
The combined fit has nine free parameters. Following the procedure
described in Ref.~\cite{PZ}, to each systematic errors quoted in the 
experimental paper is assigned a free systematic parameters which 
is then minimized in the fit along with parameters. 
As for the single-$x_\pom$ fits, 
only data for which  $\mx^2\ge 4 \, \mbox{GeV}^2$ 
and $Q^2 \ge 8.5 \, \mbox{GeV}^2$  are included in the fit.
The latter has a appreciable sensitivity on the scale $Q_0^2$
due to the relative stiffness of the initial condition in eq.~(\ref{ic}).
The choice of $Q_0^2$ is then optimized performing a scan which 
gives the best $\chi^2$ value for $Q_0^2=2.3 \, \mbox{GeV}^2$. 
The partial results for the $\chi^2$ function in the various $x_\pom$ bin are reported 
in Table \ref{results_global_xpom_fits}. 
\begin{table}[t]
\caption{Combined $x_\pom$ fit result.}
\begin{center}
\lineup
\begin{tabular}[t]{*{4}{l}}
\br
$x_\pom$ & $\chi^2$  & fitted  & total  \\
         &           & points  & points \\
\mr  
0.0003  &  3.1 &  3 & 12 \\
0.0010  & 26.2 & 19 & 40 \\
0.0030  & 38.9 & 39 & 59 \\
0.0100  & 58.6 & 61 & 80 \\
0.0300  & 36.4 & 68 & 85 \\
\br
\label{results_global_xpom_fits}
\end{tabular}
\end{center}
\end{table}
The best fit returns a $\chi^2=166$ for 182 degrees of freedom
which is of comparable quality as the one presented in Ref.~\cite{H1LRG06}. 
The initial condition, eq.~(\ref{ic2}), allows the singlet and gluon 
normalization, $A_q$ and $A_g$ respectively, to have a different power behaviour. 
It is therefore interesting to notice that if the condition $A_{q,1}=A_{g,1}$
is imposed, this result in a global $\chi^2=171$ for 183 degree of freedom. 
If one further neglects the $x_\pom$-dependence of $B_q$ and $C_q$ by setting 
$B_{q,1}=C_{q,1}=0$ the $\chi^2$ increases to 188 units for 185 degree of freedom. 
This is an \textsl{a posteriori} confirmation that not only diffractive parton distributions 
change their magnitude versus $x_\pom$ but also that a modulation in their $\beta$-shape
(for the singlet, in this case) is necessary to better fit the data. 
\section{Conclusions}
\noindent\\
We have outlined a new method to extract diffractive parton distributions
inspired by the factorization theorem for diffractive DIS.  
From a series of pQCD fits at fixed $x_\pom$ we infere the dependence of 
parameters on such a variable and this allows us to construct 
a generalized initial condition without assuming neither proton vertex factorization 
nor the existence of a series of Regge trajectories. 
Although the quality of the resulting fit 
gives a $\chi^2$ / d.o.f close to unity, as the Regge based pQCD
fit~\cite{H1LRG06}, 
the new procedure treats the non-perturbative $x_\pom$-dependence of the 
cross-sections in a controlled and less model dependent way.
This feature will allow 
further factorization test to be performed with the
improved precision of published~\cite{H1FPS} and forthcoming
data~\cite{H1LRG,H1VFPS}.


\section*{References}
\begin{thebibliography}{99}
\bibitem{DDISfactorization} J.~C.~Collins,~\textsl{Phys.~Rev.~}~\textbf{D57} (1998) 3051.
\bibitem{GTV} M.~Grazzini, L.~Trentadue, G.~Veneziano,~\textsl{Nucl.~Phys.~}~\textbf{B519} (1998) 394.
\bibitem{TV} L.~Trentadue, G.~Veneziano,~\textsl{Phys.~Lett.~}~\textbf{B323} (1994) 201.
\bibitem{extM} G.~Camici, M.~Grazzini, L.~Trentadue,  \textsl{Phys.~Lett.~}\textbf{B439} (1998) 382.
\bibitem{H1LRGdiffdijet} H1 Collaboration (A. Aktas et al.)~\textsl{JHEP} (2007) 0710:042. 
\bibitem{ZEUSdifdijet} ZEUS Collaboration (S. Chekanov et al.)~\textsl{Eur.~Phys.~J.~}\textbf{C52} (2007)813.
\bibitem{ddijet_photo_H1} H1 Collaboration (F.D. Aaron et al.)~\textsl{Eur.~Phys.~J.~}\textbf{C70} (2010) 15. 
\bibitem{ddijet_photo_ZEUS} ZEUS Collaboration (S. Chekanov et al.)~\textsl{Eur.~Phys.~J.~}\textbf{C55} (2008) 177.  
\bibitem{H1FLD} H1 Collaboration (F.D. Aaron et al.) e-Print: \texttt{arXiv}:1107.3420.
\bibitem{H1LRG06} H1 Collaboration (A. Aktas et al.)~\textsl{Eur.~Phys.~J.~}~\textbf{C48} (2006) 715.  
\bibitem{CT} F.~A.~Ceccopieri, L.~Trentadue, \textsl{Phys.~Lett.~}~\textbf{B655} (2007) 15.
\bibitem{Zeus_combo} ZEUS Collaboration (S. Chekanov et al.) ~\textsl{Nucl.~Phys.~}\textbf{B831} (2010) 1.
\bibitem{recent_global_fit} S.~Taheri Monfared, Ali N.~Khorramian, S.~Atashbar Tehrani, 
e-Print: \texttt{arXiv}:1109.0912. 
\bibitem{QCDNUM17} M.~Botje,~\textsl{Comput.~Phys.~Commun.~} \textbf{182} (2011) 490. 
\bibitem{PZ} C.~Pascaud and F.~Zomer, LAL-95-05.
\bibitem{H1FPS} H1 Collaboration (F.D. Aaron et al.)~\textsl{Eur.~Phys.~J.~}\textbf{C71} (2011) 1578. 
\bibitem{H1LRG} H1 Coll., 
[H1prelim-10-011].
\bibitem{H1VFPS} H1 Coll., 
[H1prelim-10-014], T.~Hreus, \textsl{PoS DIS2010} (2010) 068. 
\end{thebibliography}
\end{document}